\documentstyle[12pt,epsfig]{article}
\begin{document}
{\noindent \it To be published in  Physica A {\bf 270}, 222 (1999)}

\begin{center}
{\LARGE \bf The Hamming Distance in the Minority Game}

\vskip 0.6cm
{\large R.D'hulst \footnote{corresponding author: Rene.DHulst@brunel.ac.uk} and G.J. Rodgers}
\vskip 0.4cm

{\it Department of Mathematics and Statistics, Brunel University,\\
Uxbridge, Middlesex, UB8 3PH, UK.}

\end{center}

{\noindent \large Abstract}
\vskip 0.4cm

We investigate different versions of the minority game, a toy model for agents buying and selling a commodity. The Hamming distance between the strategies used by agents to take decisions is introduced as an analytical tool to determine several properties of these models. The success rate of the agents in an adaptive version of the game is compared with the rate from a stochastic version. It is shown numerically and analytically that the adaptive process is inefficient, increasing the success rate of the unused strategies while decreasing the success rate of the strategies used by the agents. The agents do not do as well as if they were forced to use only one strategy permanently. A version of the game in which the agents strategies evolve is also analyzed using the notion of distance. The agents evolve into a state in which they are all using one strategy, which is again the state that yields the maximum success rate.


\newpage
\setlength{\unitlength}{1cm}
\section{Introduction}
\label{sec:introduction}

\indent\indent A market is a network of interactions between agents buying and selling a commodity \cite{def market}. On the one hand, the scarcity or the abundance in the commodity determines its price,  while on the other hand, the price of the commodity modifies the supply and the demand equilibrium. Consequently, the variation of the price of the commodity reflects the dynamic of the market.

The internal dynamic of a market is the result of the unrelated actions of the agents buying and selling. At any one moment in time, agents are either in the group of sellers or buyers, and agents can expect to profit if they  are in the smaller group. Less sellers implies a higher price and a higher profit for the suppliers while if the number of buyers is small, the sellers will be forced to offer low prices to sell the commodity. This is a rough picture of how the internal dynamic distributes the profits amongst the agents. 

Obviously, the agents need to analyze the market's history to decide on their  future action. This history is included in the market prices or in the volume of the exchanges. Hence, every agent has a strategy that uses the past prices of a commodity to predict the future price.  In the course of time, the agents have to modify their strategies or their confidence in their different strategies to adapt themselves to the market changes. The profit of an agent will be determined by the success rate of his strategy and the efficiency with which he adapts to market changes. 

The minority game model is a very simple model that mimics the internal dynamic of the exchange of one commodity. The agents are allowed to buy or to sell a commodity at each time step and those choosing the smaller group among buyers and sellers are rewarded a point. The record of which group won for the past time steps constitutes the history of the system. Agents strive to maximise the number of points they are awarded. No attempt is made to model any external factors that influence the market.

In the model, the agents have strategies at their disposal to analyse the history of the system and to take a decision for the next time step. In this paper, we introduce the notion of Hamming distance between strategies as an analytical tool to study different versions of the minority game. 

In Sec. \ref{sec:themodel}, the basic features of the minority game model are introduced with the notion of strategy and Hamming distance between strategies. In Sec. \ref{sec:probtowin}, the success rate of the agents in an adaptive version of the model is investigated. The difference between the success rate of a strategy and the success rate of an agent is outlined. Another non-adaptive version of the model is also introduced to question the relevance of the adaptive process. It is shown that if agents are allowed to choose between strategies they do not do as well as if they were forced to use only one strategy permanently. This result is confirmed in Sec. \ref{sec:johnson} where an evolutionary version of the minority game model \cite{johnson98-2} is investigated. It is shown that most of the agents evolve spontaneously towards a state in which they are only using one strategy.

\section{The model}
\label{sec:themodel}

\indent\indent The models we consider are different versions of the bar-attendance problem introduced by Arthur \cite{arthur94} and simplified into a minority game by Challet and Zhang \cite{challet97,challet98}. Basic to the minority game is an odd number $N$ of agents having to choose between two sides, $0$ or $1$, at each time step. An agent wins if he falls into the minority side. The record of which side was the winning side for the last $m$ time steps constitutes the history of the system. For a given $m$, there are $2^m$ different histories. The 8 different histories for $m=3$ are listed in the first column of table 1.

Every agent makes a decision for the next time step according to the history of the system. To be able to play, an agent must have a strategy that allows him to make a decision for any of the different histories. The second and the third columns of table 1 list two possible sets of decisions  $\sigma$  and $\sigma'$, that we will call strategies.

If we compare  $\sigma$ and $\sigma'$ component by component, we see that for some histories they make the same prediction and for others they make opposite predictions. In this example, the decisions only differ when the history is 000, 011 or 110. The normalized Hamming distance between strategies $\sigma$ and $\sigma'$, 

\begin{equation}
d = {1\over 2^m} \sum_{i=1}^{2^m} | \sigma_i - \sigma'_i |
\end{equation}
is a measure of the difference between the two strategies. In the example of table 1, $d=3/8$. It corresponds to a time average of the different predictions of the two strategies, assuming that all the histories are equally likely to occur. Under this condition, $d$ is the probability that an agent who always uses strategy $\sigma$ will take the opposite decision to an agent who always uses strategy $\sigma'$. 

By definition, the distance is a number ranging from 0 to 1. For a given $m$, the different distances available are

\begin{equation}
d_n = {n\over 2^m} , \qquad \hbox{for }n=0,1,2,..,2^m.
\end{equation}
If we choose two strategies at random, 

\begin{equation}
P (d) = {2^m!\over 2^{2^m} (2^md)! (2^m(1-d))!}
\end{equation}
is the probability that the distance between these two strategies is $d$. We refer to $P(d)$ as the distance distribution. This is a binomial distribution with

\begin{equation}
\overline{d} = {1\over 2} 
\end{equation}
for the mean value of $d$. The variance of the distribution is $\sigma^2 (d) = 2^{-(m+2)}$. In what follows, we investigate different versions of the minority game, to show the relevance of the notion of distance as an analytical tool. 

\section{Success rates}
\label{sec:probtowin}

\indent\indent We first investigate the model introduced by Challet and Zhang \cite{challet97}. Each agent has a fixed set of $s\ge 1$ strategies at his disposal. These strategies are chosen at random from all the different possible strategies, multiple choices of the same strategy are allowed. At each time step, every agent chooses one of his strategies to predict which side will be the minority side. The winning agents are awarded a point, while the points held by the losing agents do not change. In addition to giving points to the agents, the strategies also earn points themselves. A strategy earns a point each time it has forecast the correct winning side. The points awarded to strategies, called virtual points, allow an agent to measure the efficiency of all his strategies, irrespective of whether they were used or not. This measure of the success rate of the different strategies helps the agents to choose their future strategies. 

We distinguish between a determinisitic and a stochastic version for this model, according to the way an agent chooses his playing strategy. In the former, the agents pick the strategy with the highest number of virtual points while in the latter, they simply choose a strategy at random.

There are two different issues to consider, firstly, that each strategy possesses an opposite. Consequently, half of the strategies are potentially winning and half of the strategies are potentially losing, irrespective of the number of agents playing or of the history of the previous days. Secondly, that there are $N=2k+1$ agents and at most $k$ of them can win at each time step. Hence the probability of picking a winning strategy at random in the strategy space is exactly equal to $1/2$ but the average probability that an agent wins is at most equal to $k/(2k+1)<1/2$. These two considerations and  the fact that an agent does not have access to all the strategies means that the success rate of an agent is different from the success rate of a strategy. The difference between those two success rates is a measure of the ability of an agent to choose a performing strategy.  We will now investigate the average success rates of the agents and their strategies for both versions of the model.

First of all, we choose a reference strategy: for the stochastic version, any strategy is suitable, but for the deterministic version, we choose the strategy with most virtual points \cite{johnson98-1}. Then, we assign an integer rank to all the other strategies: for the stochastic version, the rank $r$ is given to all the strategies at distance $d_r = r/2^m$ from the reference one. For the deterministic version, we give rank 0 to the strategy with the most virtual points and rank $r$ to the strategy with the ($r$+1){\raise 1ex\hbox{\small th}} largest number of virtual points. Strategies at the same distance, or with the same number of virtual points are given the same rank. We will assume that a strategy that is performing well is not very different from the best strategy, that is, the one with the most virtual points. Hence, we assume that the strategies of rank $r$ in the deterministic version are also the strategies at distance $d_r = r/2^m$ from the best one. The average number of strategies at rank $r$ is simply $P(d_r)$. As a result of these assumptions, we can estimate $N_r$, the average number of agents using a strategy of rank $r$ in the next round of decisions. In the stochastic model, 

\begin{equation}
N^{(stoch)}_r = N P (d_r).
\label{eq:nr stochastic}
\end{equation}
In the deterministic version, an agent uses a strategy of the $r${\raise 1ex\hbox{\small th}} rank only if he has no strategy of a lower rank. The average number of agents using the best strategy, $N_0^{(det)}$, is

\begin{equation}
N_0^{(det)} =  N \left(1 - \left( 1-{1\over 2^{2^m}}\right)^s\right),
\label{eq:n0 deterministic}
\end{equation}
while the average number of agents using a strategy of rank $r>0$, $N_r^{(det)}$, is

\begin{equation}
N_r^{(det)} = N \sum_{i=0}^{s-1} {s!\over i! (s-i)!} (P(d_r))^{s-i} \bigl( 1-\sum_{j=0}^r P(d_j)\bigr)^i
\end{equation}
also equal to

\begin{equation}
N_r^{(det)}= N( \bigl( 1-\sum_{j=0}^{r-1} P(d_j)\bigr)^s - \bigl( 1-\sum_{j=0}^r P(d_j)\bigr)^s).
\label{eq:nr deterministic}
\end{equation}
The average distance to the reference strategy is equal to

\begin{equation}
\overline{d} = {1\over N} \sum_{r=0}^{2^m} N_r d_r,
\end{equation}
with $N_r$ taking the form of Eq. (\ref{eq:nr stochastic}) or (\ref{eq:n0 deterministic}) and (\ref{eq:nr deterministic}) for the stochastic and deterministic versions of the model respectively. Notice that $\overline{d}\le 0.5$, with the equality holding only for the stochastic version of the model. 

Suppose now that there are $N=5\ (k=2)$ agents playing, numbered from 1 to 5. $d_{ij}$ is the distance between the strategies used by agents $i$ and $j$. An agent choosing one side wins if at least $k+1=3$ other agents choose the other side. The success rate of agent 1, $\tau_{ag}^{(1)}$, defined as his probability to win a point per time step, is equal to 

\begin{eqnarray}
\tau_{ag}^{(1)}= d_{12}d_{13}d_{14}d_{15}&+&d_{12}d_{13}d_{14}(1-d_{15})
+ d_{12}d_{13}(1-d_{14})d_{15}\nonumber\\ 
&+&d_{12}(1-d_{13})d_{14}d_{15}
+ (1-d_{12})d_{13}d_{14}d_{15}.
\label{eq:probability to win1}
\end{eqnarray}
The first term on the right hand side is the probability that agent 1 chooses one side and the other 4 agents the other side; the 4 other terms on the right hand side are the probabilities that only one agent chooses the same side as agent 1. For any other combination of choices agent 1 would lose. This can be generalized to any number of agents so that the average success rate for an agent can be expressed as 

\begin{equation}
\tau_{ag} = \sum_{i=0}^{N/2-3/2} {(N-1)!\over i!(N-1-i)!} \overline{d}^{N-1-i} (1-\overline{d})^i
\label{eq:successagents}
\end{equation}
where we have averaged over all the distances. For the deterministic version of the model, $\overline{d}$ is a function of $s$, but for the stochastic version, 

\begin{eqnarray}
\tau_{ag}^{(stoch)} &=& \Bigl( 2^{N-2} - {(N-1)!\over 2 \bigl({N-1\over 
2}\bigr)!^2}\Bigr) {1\over 2^{N-1}}\nonumber\\
&\simeq& {1\over 2} - exp\biggl(1-{1\over 2}\ ln(2\pi N) 
+ N ln \Bigl({N\over N+1}\Bigr)\biggr)
\label{eq:probability to win2}
\end{eqnarray}
where we have put $\overline{d}=1/2$. For the stochastic version of the model, all the agents are identical and $\tau_{ag}^{(stoch)}$ is an estimation of the success rate of the agents. The numerical and analytical values of $\tau_{ag}^{(stoch)}$ are compared in Fig.1a (two lower curves) for $N=1001$ and $m=5$. The agreement is excellent. Notice that the success rate of the agents is independent of $s$.

For the deterministic version, Eq. (\ref{eq:successagents}) gives the success rate of the reference agent, $\tau_{ag}^{0}$, which goes to zero very fast as a function of $s$. According to our classification, the success rate of an agent of rank $r$ is

\begin{equation}
\tau_{ag}^{r} = (1-d_r) \tau_{ag}^{0}  + d_r (1 - \tau_{ag}^{0}).
\end{equation}
which is approximately equal to $d_r$ for $s \ge 2$. The average success rate of any agent in the deterministic version is then

\begin{equation}
\tau_{ag}^{(det)} = {1\over N}\sum_{r=0}^{2^m} N_r^{(det)} \tau_{ag}^{r}.
\end{equation}
This expression is compared to numerical results in Fig.1b (two lower curves) for $N=1001$ and $m=5$. The agreement is qualitatively good. The quantitative discrepancies probably originate from our assumption that there is only one strategy with the highest number of virtual points. In fact, we also showed that in this case, the success rate of this strategy is approximately 0. Hence, most of the time there is more than one strategy with the highest number of virtual points. 

Finally, we consider the success rates of the strategies in both versions of the model. When a strategy is used, it is in competition with one strategy from each of the other agents. Consequently, it has the same average success rate as an agent. On the other hand, when a strategy is not used, it earns points only if it can predict the winning side. In the stochastic version, each side is as likely to be the winning side. Moreover, there are as many strategies suggesting one side as strategies suggesting the other. Hence, the average success rate for a strategy in the stochastic version of the model, $\tau_{st}^{(stoch)}$, is

\begin{equation}
\tau_{st}^{(stoch)} = {0.5(s-1) + \tau_{ag}\over s}.
\end{equation} 
In Fig.1a this analytical result is compared with numerical simulations of the system for $N=1001$ and $m=5$. There is qualitative agreement between the two curves. 

For the deterministic version, the strategies that are not used have a greater average success rate than 0.5: they are not used because they differ from what we called the best strategy, the one with the most virtual points. But this strategy loses most of the time, as its very low success rate suggests. The more different a strategy is from the one with largest number of vitual points, the more likely it is to win. We can express the success rate of the unused strategies $\tau_{st}^{nu}$ as

\begin{equation}
\tau_{st}^{nu} = {1\over N} \sum_{r=0}^{2^m} N_r^{(det)} {\sum_{i=r}^{2^m} d_i P(d_i)\over \sum_{i=r}^{2^m} P(d_i)}.
\label{taustnu}
\end{equation}
The average success rate of a strategy for the deterministic version is then equal to

\begin{equation}
\tau_{st}^{(det)} = {\tau_{st}^{nu}(s-1) + \tau_{ag}^{(det)}\over s},
\label{eq:taustdet}
\end{equation} 
because at each time step, one strategy is used and $s-1$ are not. The numerical and analytical results for $\tau_{st}^{(det)}$ are also reported in Fig.1b (two upper curves) for the same set of values. The two curves are in qualitative agreement.

If we compare in both models the success rate of the agents and the success rate of their strategies, the agents seem to make very poor choices. In a set of strategies of a certain average success rate, the agents always pick strategies with a success rate lower than the average. In the stochastic model this occurs because the unused strategies have a success rate of 0.5 and there are more unused strategies for greater $s$. For the deterministic version, the adaptive process is such that the agents try to concentrate around the same points in the strategy space. Hence, they are all adaptating in the same way, making the same predictions and taking the same decisions. This process is only restricted by the part of the phase space available to the agents. Consequently, as $s$ increases their performence deteriorates. Conversely, their adaptive process increases the success rate of the unused strategies instead of increasing the success rates of the agents. Obviously the problem with the adaptive process is that every agent is using the same strategy to adapt. To improve the model, an evolutionary process allowing for selection of efficient adaptive processes is required. In the next section, we investigate a version of the minority game model in which the strategies of the agents are allowed to evolve. 

\section{The minority game with evolving agents}
\label{sec:johnson}

\indent\indent The second version of the minority game we investigate is a model introduced by Johnson {\it et al.} \cite{johnson98-2}. Every agent $i$ has one strategy at his disposal. Some agents might share the same strategy.  At each time step, each agent follows the prediction of his strategy with a probability $p_i$, where $i$ denotes the agent number $i = 1,...,N$. The agent takes the opposite decision with a probability $1-p_i$. Each time an agent chooses the minority side, he wins and earns a point. Otherwise, he loses and loses a point. The worst performing agents are allowed to change their probability $p_i$ before the next time step commences. An agent scoring less than a given lower value $L(<0)$ chooses a new probability $p_i$ at random from the range $\lbrack p_i-R/2, p_i+R/2\rbrack$. Reflective boundary conditions are used, but our results do not depend on the particular choice of boundary conditions. The agent's score is then reset to 0. The free parameter $R$ allows one to adjust the persistence with which an agent sticks to an unsuccessful approach. 

We define $Q(p)dp$ as the probability that an agent picked at random has an associated probability from the range $\lbrack p-dp/2, p+dp/2\rbrack$. Hence, $Q(p)$ is the probability distribution of the probabilities $p$. We estimate the probability distribution $Q(p)$ using the notion of distance.

When an agent chooses a new strategy he is said to mutate. As the new strategy is choosen from the range $\lbrack p_i-R/2, p_i+R/2\rbrack$, the time evolution of $Q(p)$ is given by

\begin{eqnarray}
\nonumber
Q (p , t + dt) &=& Q (p , t) - S (p , t) dt \\
&+& {dt\over 2} (S (p + \Delta p , t) + S (p - \Delta p , t))
\end{eqnarray}
where $S(p,t)$ is the probability that any agent with probability $p$ will mutate at time $t$. This can be rewritten as

\begin{equation}
{\partial Q\over \partial t} = {\Delta p^2\over 2} {\partial^2 S\over \partial p^2}
\label{eq:diffusion}
\end{equation}
with the normalization condition

\begin{equation}
\int_0^1 dp Q (p) = 1.
\label{eq:normal}
\end{equation}
As $\Delta p$ is a measure of the variation of the probability associated with an agent when this agent mutates, we write $\Delta p \simeq R/4$. If $\Delta t_m (p)$ is the time between two mutations for an agent associated with a probability $p$, the probability of mutation of this agent is $1/\Delta t_m (p)$ per time step. Hence, 

\begin{equation}
S (p,t) = {Q (p,t)\over \Delta t_m (p)}
\end{equation}
on average. We call $\tau (p)$ the success rate of an agent with associated probability $p$, supposing that this success rate depends only on $p$. This means that $\tau (p)<1/2$. On average, the number of points earned by an agent, $C (p,t)$, evolves as

\begin{equation}
C (p,t) = C (p,t_0) - \left({1\over 2} - \tau (p) \right) (t-t_0).
\end{equation}
Applying this equation to $t-t_0 = \Delta t_m (p)$, 

\begin{equation}
\Delta t_m (p) = -{L\over {1\over 2}-\tau (p)}
\end{equation}
because an agent mutates when $C(p,t)=L (<0)$ and, after mutation, his points are reset to $C (p,t_0) = 0$. Hence,

\begin{equation}
{\partial Q\over \partial t} = {R^2\over 32L} {\partial^2\over \partial p^2}\left( \left(\tau (p) - {1\over 2}\right)Q(p)\right).
\label{eq:evolution of Q(p)}
\end{equation}
Eq. (\ref{eq:evolution of Q(p)}) is a diffusion equation for the probability distribution in the space of $p$. The diffusion coefficient, $(\tau - 0.5)R^2/32 L$, is a function of the position $p$. The normalization condition of Eq. (\ref{eq:normal}) corresponds to a mass conservation law for $Q(p)$. Integrating Eq. (\ref{eq:evolution of Q(p)}) with respect to $p$ gives

\begin{equation}
{\partial \over \partial p} \left(\left({1\over 2} - \tau (p) \right)Q(p)\right) = 0,
\end{equation}
which can be rewritten as

\begin{equation}
Q (p) =  {N_0 \over {1\over 2} - \tau (p)}
\label{eq:equilibrium Q(p)}
\end{equation}
where $N_0$ is determined by the normalization of $Q(p)$.

We need to estimate the success rate of an agent as a function of $p$. Consider first any agent $i$. He is characterized by his strategy $\sigma_i$ and his probability $p_i$, that is, by the pair $(\sigma_i ,p_i)$. At each time step, he uses $\sigma_i$ with a probability $p_i$ and $1-\sigma_i$ with a probability $1-p_i$. On average, this agent uses an effective strategy defined as

\begin{equation}
\sigma_i^{eff} = p_i \sigma_i + (1-p_i ) (1-\sigma_i ).
\end{equation}
Switching to the components, to each $1$ in $\sigma_i$ corresponds a $p_i$ in $\sigma_i^{eff}$, while to each $0$ in $\sigma_i$ corresponds a $1-p_i$ in $\sigma_i^{eff}$. Consider now two agents $i$ and $j$ characterized by pairs $(\sigma_i ,p_i)$ and $(\sigma_j ,p_j)$ respectively. Comparing the effective strategies of these two agents component by component, there are 4 possible different realisations of the components, as presented in table 2. In this table, $(\sigma_i^{eff})_k$ is component $k$ of the effective strategy of agent $i$.
If $d$ is the distance between $\sigma_i$ and $\sigma_j$, $d_{eff}$, the distance between the effective strategies of agents $i$ and $j$, is

\begin{equation}
d_{eff} = d |1 - (p_i + p_j )| + (1-d) |p_i - p_j |.
\end{equation}
We call $d_{eff}$ the effective distance between $(\sigma_i ,p_i)$ and $(\sigma_j ,p_j)$. To find the average effective distance between agent $i$ and the other agents, two averaging processes must be performed. The first one is over $d$, using the distance distribution $P  (d)$, and gives

\begin{equation}
\langle d_{eff}\rangle_{P(d)}  = {1\over 2} ( |1 - (p_i + p_j )| + |p_i - p_j |).
\end{equation}
The second average is over the probability distribution $Q(p_j)$ and gives

\begin{equation}
\overline{d}_{eff} (p) = {1\over 2} \int_0^1 dp_j Q(p_j ) ( |1 - (p + p_j )| + |p - p_j |)
\label{eq:average effective distance}
\end{equation}
where we drop the subscript $i$ from now on. The success rate as a function of $p$ is then given by

\begin{equation}
\tau (p) = \sum_{i=0}^{N/2-3/2} {(N-1)!\over i!(N-1-i)!} \overline{d}_{eff}^{N-1-i} (1-\overline{d}_{eff})^i
\label{eq:evolution t(p)}
\end{equation}
using Eq. (\ref{eq:successagents}).

Equations (\ref{eq:equilibrium Q(p)}), (\ref{eq:average effective distance}) and (\ref{eq:evolution t(p)}) completely determine $\tau (p)$ and $Q (p)$. To approximate both these quantities, we assume that $Q(p)$ has a uniform distribution, that is $Q(p)=1$ for $p$ between 0 and 1. Eq. (\ref{eq:average effective distance}) gives

\begin{equation}
\overline{d}_{eff}(p) = {1\over 2} (1-2p + 2p^2 ).
\end{equation}
The average effective distance is a maximum at $p=0$ and $1$ where it is equal to 0.5. Otherwise, $\overline{d}_{eff}< 0.5$, achieves its minimum value $0.25$ at $p=0.5$. Consequently, the success rate of the agents for $p$ around 0 or 1 is higher than for $p$  around 0.5. Iterating equations (\ref{eq:equilibrium Q(p)}), (\ref{eq:average effective distance}) and (\ref{eq:evolution t(p)}) allows us to make successive approximations to $Q(p)$. An approximate solution for $N Q(p)$ after a hundred iterations is presented in Fig.2 with the results of a direct numerical simulation of the system. We repeated our iterations starting with a Gaussian distribution, an asymmetric distribution and a parabolic distribution for $Q(p)$. These all converged to the same result for the set of parameters presented at Fig.2. In fact, only 2 or 3 iterations were enough to reach a stationary distribution. As Fig.2 illustrates, the agreement between the numerical iteration of the  analytical results and the direct numerical simulation of the model is very good. 

It is interesting to note that most of the agents are associated with a probability near 1 or 0. Such agents use only one strategy most of the time, either their own or the opposite one, respectively. This conclusion compares with the conclusion of the previous section. In fact, it was shown that if the agents are allowed to choose between strategies they do not do as well as if they were forced to use only one strategy permanently. In the adaptive model of the previous section, the agents were not able to recognize this fact and performed badly. In the evolutionary model we investigated in this section, the agents are able to recognize this and most of them evolved to a state where they were using only one strategy. Hence, a generalization of this evolutionary model seems to be the solution of the problem of the adaptive model. 

\section{Conclusions}
\label{sec:conclusions}

\indent\indent We have shown how the Hamming distance between strategies is a useful tool to investigate the properties of the minority game. It is a measure of the difference between two strategies and corresponds to a time average of the different predictions of these two strategies, assuming that all the histories are likely to occur. We successfully applied this notion to compute the average success rate of the agents in different versions of the minority game. For every version, the analytical results were complemented by appropriate numerical simulations.

In an adaptive version of the model, the difference between the success rates of the used and unused strategies was emphasized. It was shown that the adaptation process is ineffective, increasing the success rate of the unused strategies instead of the success rate of the strategies used by the agents. In fact, if the agents were forced to use one strategy permanently, they would do better. To illustrate this conclusion, the adaptive model was compared to a stochastic version of the minority game. It was shown that adaptive agents having a set of $s$ strategies at their disposal perform worst than agents taking decisions according to a strategy chosen a random at each time step among the same set of strategies. Our analytical results suggest that the problem with the adaptive process is that every agent is using the same fixed mechanism to adapt. As an improvement, we suggest it would be interesting to let the method used by agents to select strategies evolve. Consequently, a version of the minority game in which the agents strategies evolve was investigated.

In the evolving version of the model, the agents have only one strategy, and its opposite, at their disposal. It was shown numerically and analytically that most of the agents spontaneously use only one of these strategies. Hence, in this version of the minority game as well as in the previous adaptive version, an agent benefits from always using the same strategy. In the adaptive version of the game, agents do not realize this, in the evolving version, the agents evolve to this state. We proposed that a simple generalization of this model will improve the efficiency of the adaptive process.

As a corollary, let us mention the work of Cavagna \cite{cavagna98}, who claims that the only relevant parameter for the memory is the value of $m$, not the information contained in the memory. In fact, all our results are obtained using only geometrical considerations, supporting this conclusion.

\newpage
{\noindent \large Table 1}

\begin{table}[h]
\centering
\begin{tabular}{|c|c|c|}
\hline
data & $\sigma$  & $\sigma'$\\
\hline
000 & 1 & 0 \\
001 & 1 & 1 \\ 
010 & 0 & 0 \\ 
011 & 0 & 1 \\
100 & 1 & 1 \\
101 & 1 & 1 \\
110 & 0 & 1 \\
111 & 0 & 0 \\
\hline
\end{tabular}
\end{table}
\vskip 3cm

{\noindent \large Table 2}

\begin{table}[h]
\centering
\begin{tabular}{|c|c|c|}
\hline
$(\sigma_i^{eff})_k$ & $(\sigma_j^{eff})_k$  & $|(\sigma_i^{eff})_k - (\sigma_j^{eff})_k |$\\
\hline
$p_i$ & $p_j$ & $|p_i - p_j |$ \\
$1 - p_i$ & $p_j$ & $|1 - (p_i + p_j )|$ \\
$p_i$ & $1 - p_j$ & $|1 - (p_i + p_j )|$ \\
$1 - p_i$ & $1 - p_j$ & $|p_i - p_j |$ \\
\hline
\end{tabular}
\end{table}

\newpage
{\noindent \large Table Captions}

\vskip 1.0cm
{\noindent \bf Table 1} --- The first column lists all the possible histories of the system for the last three time steps $(m=3)$. A strategy is a set of decisions for all the different possible histories. Two example strategies $\sigma$ and $\sigma'$ are shown in the second and third columns.

\vskip 1.0cm
{\noindent \bf Table 2} --- The first two columns list the 4 possible realisations of the components of the effective strategies of two agents characterized by pairs $(\sigma_i ,p_i)$ and $(\sigma_j ,p_j)$. $(\sigma_i^{eff})_k$ is component $k$ of the effective strategy of agent $i$. The last column gives the different possible contributions to the calculation of the effective distance.

\newpage
{\noindent \large Figure Captions}

\vskip 1.0cm
{\noindent \bf Figure 1} --- Analytical results (line) and numerical simulations (line and crosses) for the success rates of the agents and their strategy for (a) the stochastic and (b) the deterministic versions of the model; the two upper curves are the success rates of the strategies and the two lower curves are the success rates of the agents, for both (a) and (b). The choice of the parameters is $N=1001$ and $m=5$. 40 simulations were done over 500 time steps.

\vskip 1.0cm
{\noindent \bf Figure 2} --- Comparison of the analytical (line) and numerical (line and crosses) results for $NQ(p)$, the average number of agents with a probability $p$, for $N=101$, $m=3$, $L=-4$ and $R=0.2$. The results of the direct numerical simulation are obtained after 200 simulations of $10^6$ time steps.
\newpage

\begin{picture}(21,27)(3,-5)
\epsfig{file=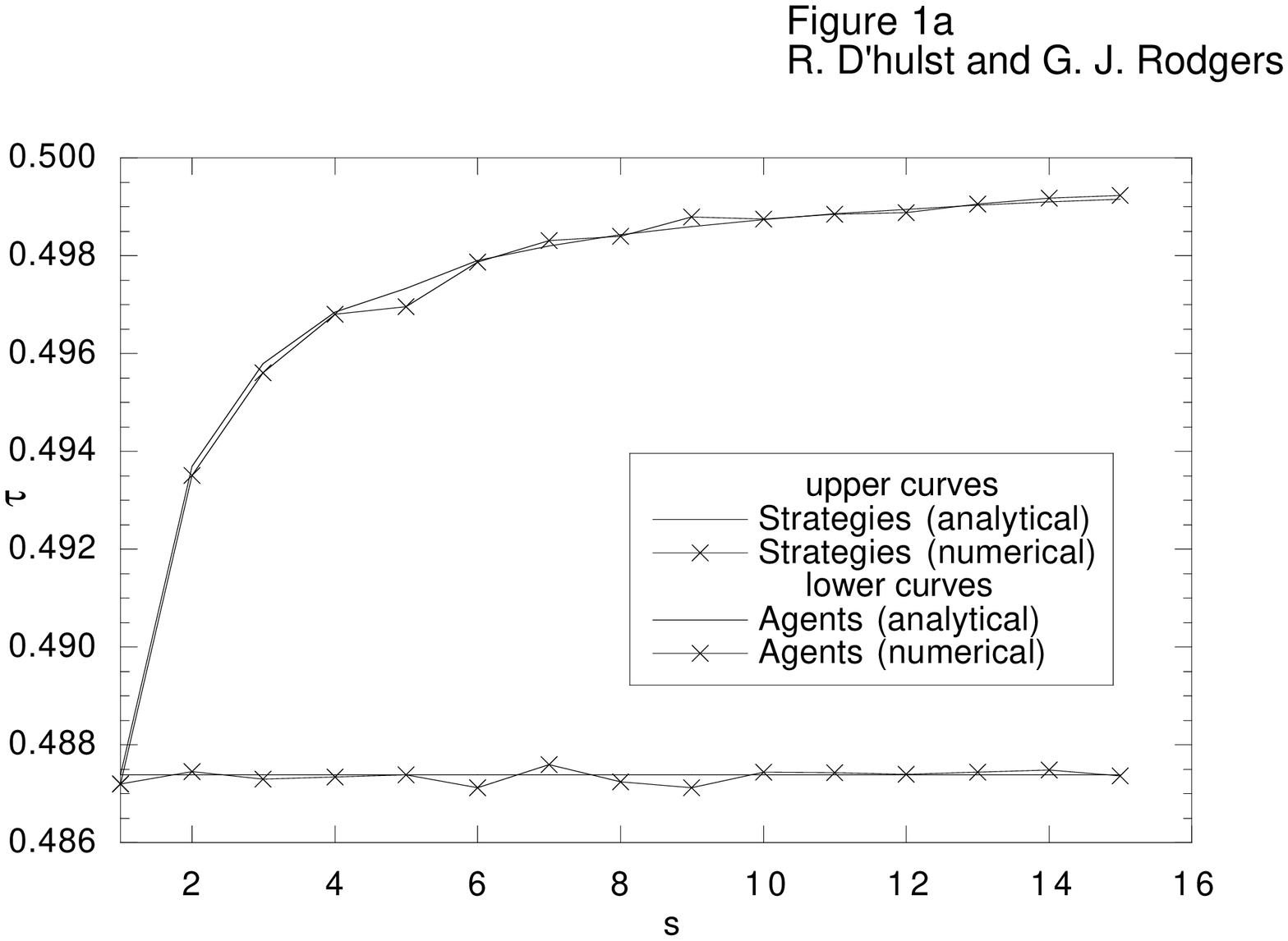,angle=90}
\end{picture}
\newpage

\begin{picture}(21,27)(3,-5)
\epsfig{file=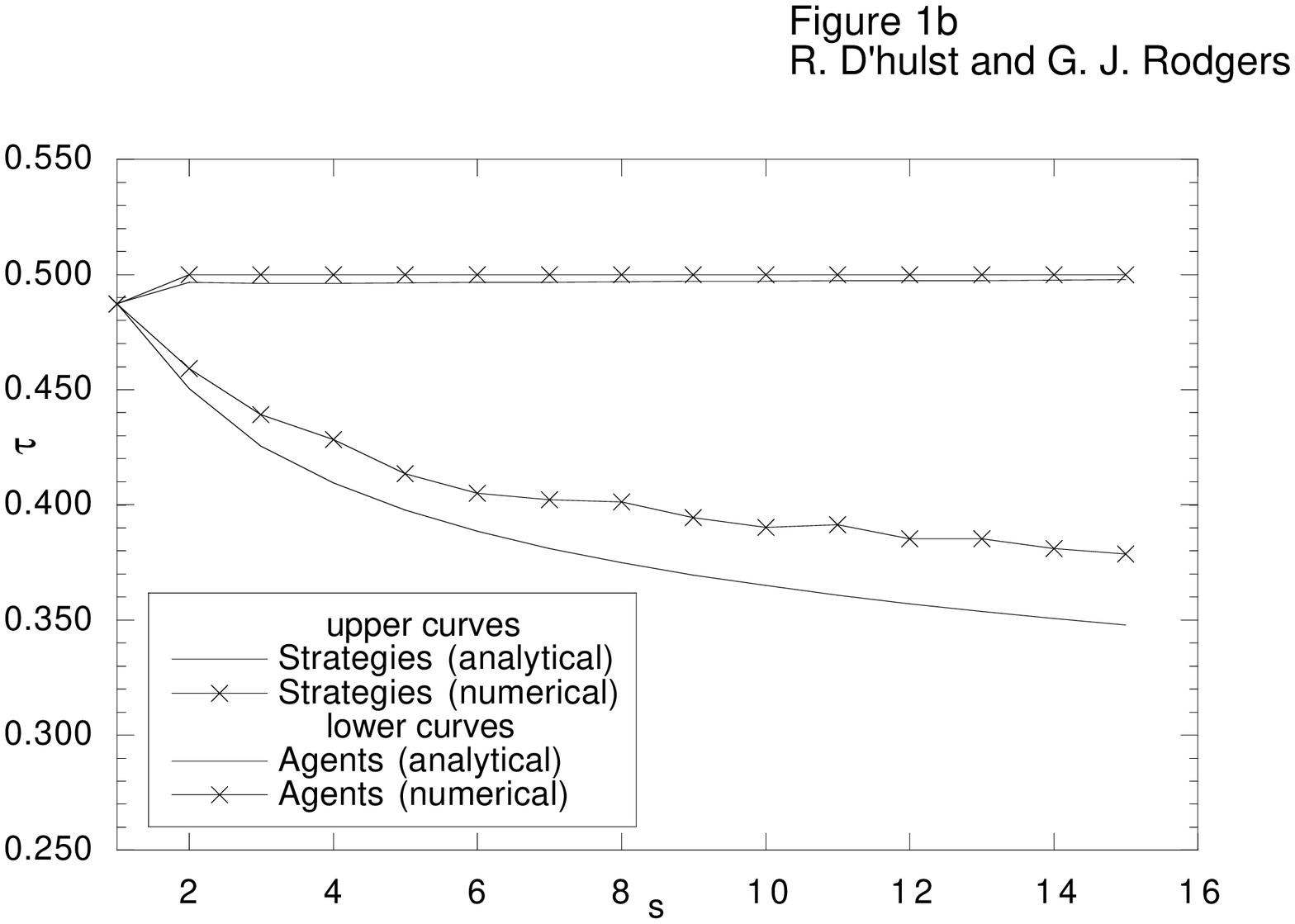,angle=90}
\end{picture}
\newpage

\begin{picture}(21,27)(3,-5)
\epsfig{file=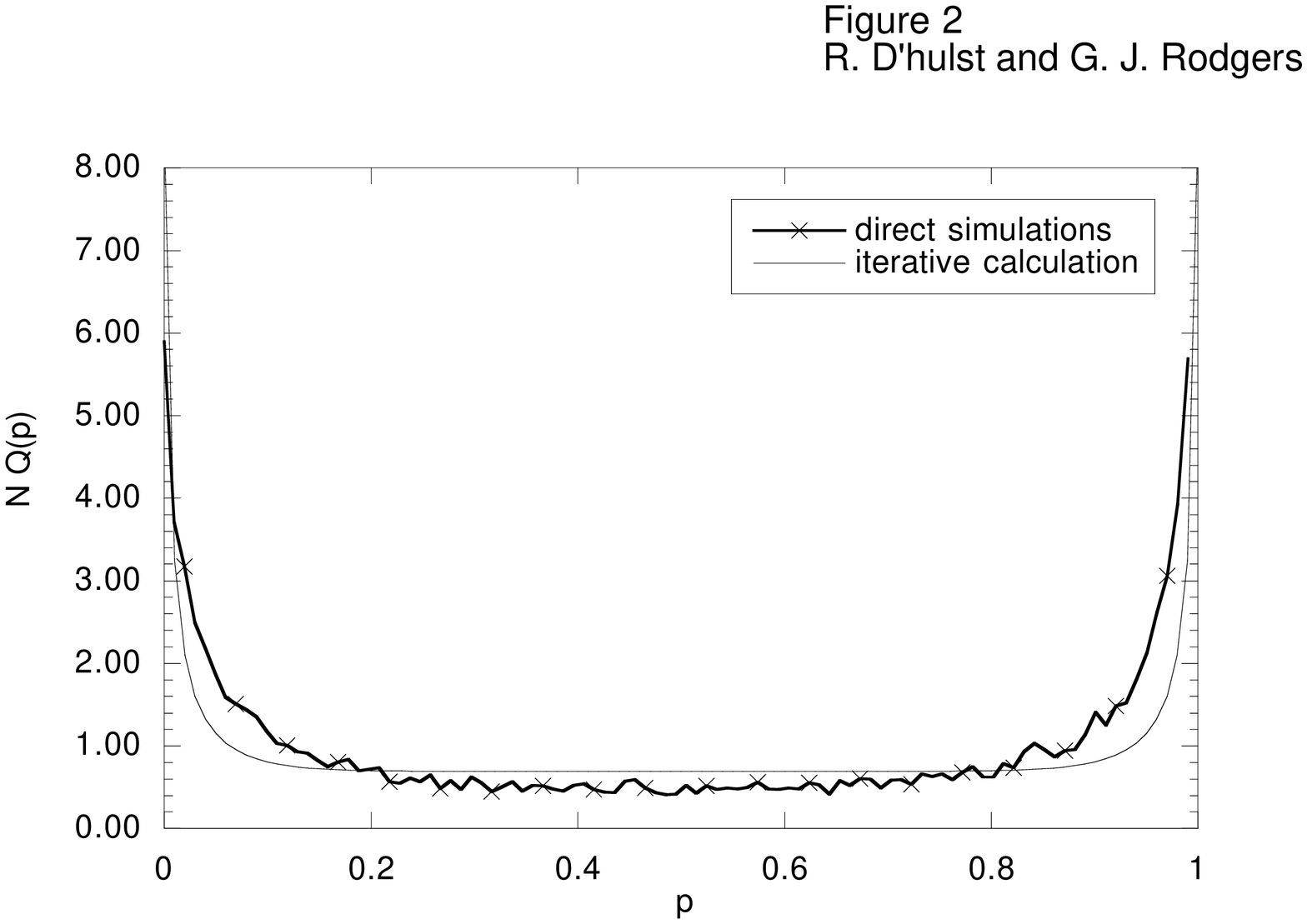,angle=90}
\end{picture}

\end{document}